# Comments on the "Regional climate variability driven by foehn winds in the McMurdo Dry Valleys, Antarctica"

Krzysztof Sienicki
*Chair of Theoretical Physics of Naturally Intelligent Systems, Topolowa 19*
*05-807 Podkowa Leśna, Poland, EU*
(24 June, 2013)

Recently Speirs *et al*. (Speirs et al. 2012) in a paper entitled *Regional climate variability driven by foehn winds in the McMurdo Dry Valleys, Antarctica* reported the studies of the relationship of "The intra- and interannual frequency and intensity of foehn events" at the McMurdo Dry Valleys and the region's climate.

Despite a substantial effort to obtain and present the results the authors work is fundamentally flawed. The authors using a mixture of poorly defined terms like "Foehn day" and "significant correlations" are trying to find a relationship between the equally poorly defined Foehn wind events at the McMurdo Dry Valleys and El Niño Southern Oscillation (ENSO) and the Southern Annular Mode (SAM).

The main objection to Speirs' *at al*. work arises from the lack of analyses of the probability distribution functions of underlying processes leading to wind formation of which velocities are measured by automated weather stations and reported in the paper.

A standard (fundamental) assumption of statistical mechanics is that physical quantities of the system under consideration are extensive variables, meaning that the physical property of the system depends on the size of the system. (Clusel and Bertin) The paradigmatic example of such a system is Brownian motion, where the sum of a great number of small motions is Gaussian distributed and growing with time variance. Thus the systems with a short range and uncorrelated interactions are usually described by extensive variables controlled by the central limit theorem.

However, we know that in nature one can observe phenomenas of much greater spatiotemporal distance than Brownian motion. One would suggest that the wind phenomena may represent a paradigmatic example of a system of which one variable like near surface wind velocity is a non extensive variable due to long range interactions. It was shown that in the case of non extensive variables ($\chi^2$-superstatistics (Tsallis statistics)) new power-law and Pareto distributions are indeed observed (Tsallis 2009).

Since power-law distributions have different properties than Gaussian distributions it is therefore fundamentally important to investigate wind velocity distribution function before making further analysis. For example the authors in Table II present their calculations of "Correlation statistics for average seasonal foehn days (FD) and air temperatures against ENSO and SAM". However, mathematically a rigorous definition of calculating the correlation coefficient (Pearson product-moment correlation coefficient) of averages *does not exist*. Therefore the author's numbers as given in Table II represent a set of randomly calculated figures. The authors suggestion in relation to a few of these random numbers that some of them have "statistical significance at the 95% level" is erroneous since no relationship exists between correlation coefficient of averages and statistical significance. Therefore Speirs *et al*. main conclusion that "the SAM is found to significantly influence foehn wind frequency" at McMurdo Dry Valleys is unfounded.

It was shown (Sienicki 2011) that wind velocities recorded at McMurdo Dry Valley did not pass the Kolmogorov–Smirnov and Shapiro–Wilk normality tests and that the wind events are self-organized criticality. Moreover it was found that the wind events recorded at weather stations located at Taylor Glacier, Lake Bonney, Lake Hoare, Lake Fryxell and Explorers Cove are distributed according to a power-law distribution function with scaling parameters $\alpha$ smaller than 2. In such a case the average of a random variable x with probability distribution function p(x) is diverging to infinity

$$\langle x \rangle = \int_0^\infty x p(x) dx \approx \int_{x_{min}}^\infty x p(x) dx = \frac{a}{2-\alpha} [x^{-\alpha+2}]_{x_{min}}^\infty \to \infty.$$

and cannot be calculated, as the product under the integral is nonintegrable.

Therefore it is pointless to suggest or even calculate the average value of the stochastic $\alpha$-stable process with $\alpha \leq 2$. One should add that having several measurements one can calculate the sum of these measurements and divide it by a different set of figures to obtain an arithmetical average which bears no relation to the above equation.

The second objection is related to the authors' unfounded ability to distinguish and identify Foehn wind events from other wind events at the McMurdo Dry Valleys, like for example katabatic and glacier winds. How by looking at wind velocity (stochastic time series) they were only able to distinguish Foehn winds from other winds? To account for Foehn winds the authors' introduce a rather artificial measure of the Foehn day - if recorded wind velocity is emanating 5 m/s from a westerly direction then a warming of at least +1°C per hour and a decrease of relative humidity of at least 5% per hour for more than 6 h occurs. However such a definition of the Foehn wind does not account for its ergodic physical nature.

Even if one assumes that authors definition of Foehn day has some merit, one easily observes that the authors are not dealing with independent and identically distributed (i.i.d.) random variables but rather with the peak-over-threshold random variables described by generalized extreme value distribution.